\documentclass[aps,prl,twocolumn,showpacs,floatfix,superscriptaddress]{revtex4-1}
\usepackage{graphicx}
\usepackage{multirow}
\usepackage{amsmath}
\usepackage{color}
\usepackage[normalem]{ulem}

\newcommand{\R}{\mathbf{r}}

\begin{document}

\title{Optical Spectra of Solids Obtained by Time-Dependent Density-Functional Theory with 
the Jellium-with-Gap-Model Exchange-Correlation Kernel}
\author{Paolo E. Trevisanutto}
\affiliation{National Nanotechnology Laboratory (NNL), Istituto di Nanoscienze-CNR,
Via per Arnesano 16, I-73100 Lecce, Italy }
\affiliation{European Theoretical Spectroscopy Facility (ETSF)}
\author{Aleksandrs Terentjevs}
\affiliation{National Nanotechnology Laboratory (NNL), Istituto di Nanoscienze-CNR,
Via per Arnesano 16, I-73100 Lecce, Italy }
\author{Lucian A. Constantin}
\affiliation{Center for Biomolecular Nanotechnologies @UNILE, Istituto Italiano di Tecnologia,
Via Barsanti, I-73010 Arnesano, Italy}
\author{Valerio Olevano}
\affiliation{European Theoretical Spectroscopy Facility (ETSF)}
\affiliation{Institut N$\acute{e}$el, CNRS and UJF, Grenoble, France}
\author{Fabio Della Sala}
\affiliation{National Nanotechnology Laboratory (NNL), Istituto di Nanoscienze-CNR,
Via per Arnesano 16, I-73100 Lecce, Italy }
\affiliation{Center for Biomolecular Nanotechnologies @UNILE, Istituto Italiano di Tecnologia,
Via Barsanti, I-73010 Arnesano, Italy}

\date{\today}
%
%
\begin{abstract}
Within the framework of \textit{ab initio} Time Dependent-Density Functional Theory (TD-DFT), we propose a static approximation to the exchange-correlation kernel based on the \textit{jellium-with-gap} model. This kernel accounts for electron-hole interactions and it 
is able to address both strongly bound excitons
and weak excitonic effects. TD-DFT absorption spectra of several bulk materials (both semiconductor and insulators) are reproduced in very good agreement with the experiments and with a low computational cost.
\end{abstract}

\maketitle

The theoretical description of the optical properties of
 materials
by first principles calculations is one of the classical issues in solid state physics.
After photon absorption, the excitonic effects, driven by  the electron
and hole (e-h) interactions, are principal actors, rendering the \textit{ab initio}
computational description demanding.
The most successful approach is, so far,
based on Many-Body Perturbation theory: the GW self energy
accounts for electron-electron (e-e) many-body effects in the band structure calculations, whereas
the Bethe-Salpeter Equation is solved to introduce 
e-h interactions \cite{Onida_rmp}.
This accurate method has a limited applicability to large systems
due to its high computational cost.

Time-Dependent Density-Functional Theory (TD-DFT) is another theory
for the exact treatment of excited states. Similarly to ground state DFT,
its main drawback is to properly approximate the \textit{unknown} \textit{dynamic}
exchange-correlation (xc) kernel $f_{xc}$, which should account for both e-e and e-h interactions.
From the very first attempt, the Homogeneous Electron Gas (HEG) model system
had been employed in construction of xc kernels \cite{ZS,RG_theo}, and one of the
most used  approach
was derived from local density approximation (LDA) xc potential $v_{xc}$ in the \textit{static} regime, Adiabatic LDA (ALDA) \cite{ZS}.
ALDA has been successfully applied
to molecules and clusters, but it usually becomes inappropriate in solids where the improvements with respect to the Random Phase Approximation (RPA)
(with the trivial $f_{xc}=0$) are negligible.
The introduction of short-range non-locality, like in a generalized-gradient approximation (GGA) \cite{GGA} or in a non-LDA (NLDA) \cite{NLDA} approach,  does not improve with respect to LDA or RPA optical spectra.
Main reason for this failure resides in the missed ultra-long range behaviour, $ 1/|\R-\R'| $ or $ 1/q^2 $ in reciprocal space, 
absent in HEG and HEG-based kernels.
On the other hand, the introduced empirical long range contribution (LRC) kernel, $f_{xc}= \alpha/q^2$ \cite{Valerio_thesisII,LRC,Botti_MgO}, with $\alpha$ found related to the screening, correctly reproduces e-h excitonic effects
but it is limited to semiconductors and small gap insulators. 
Very recently, renewed interest in this issue
has been boosted by two new approaches. In the first one, the \emph{Bootstrap}
 kernel \cite{Bootstrap} extends the LRC kernel to a matricial form and proposes a heuristic form for the LRC weight in terms of the screening which has to be calculated self-consistently.  
This method has been successfully applied to a wide range of bulk materials. 
In the second work \cite{Meta-gga},
starting from the meta-GGA xc functional they derive an $f_{xc}$ which presents an $\alpha/q^2$ term, with $\alpha$ calculated from the meta-GGA parametrizations.
Finally, the \textit{Nanoquanta} kernel \cite{Nanoquanta,Marini}  adapted from the four-point Bethe Salpeter Equation kernel
in the TD-DFT framework, achieves high accuracy, though without much less computational cost.

%
%

In this letter, we propose a non-empirical static xc kernel based on the \textit{jellium-with-gap} model (JGM) \cite{LL,Callaway,RS},
so relying on a well defined physics.
In the spirit of DFT, this kernel is a \textit{density-functional} $f_{xc}[n](E_g)$ and it also depends on the bandgap $E_g$.
The approach is \textit{ab initio} in the same regards as the LRC, Bootstrap and Nanoquanta, once first principle calculations (like GW) are used to estimate good $E_g$.
We show that this kernel properly describes both weak excitonic effects in semiconductors as well as bound excitons in ionic insulators, without requiring any frequency-dependence.
It provides absorption spectra in good agreement with experiments.
Moreover, this approach is as computationally expensive as standard ALDA.

%
%

In linear-response TD-DFT, the central quantity is the
density-density response function $\chi(q,\omega)$ (written in reciprocal space $q$ and frequency $\omega$),
which is calculated by the Dyson equation: 
\begin{equation}
\chi^{-1}(q,\omega)= \chi_0^{-1}(q,\omega) -f_{xc}(q,\omega) - v(q),
\label{Dysoneq}
\end{equation}
where $\chi_0(q,\omega)$ is the independent-particle Kohn-Sham response function, and
$v(q)=4\pi /q^2$ is the Coulomb potential. The dielectric function $\epsilon$ is related to the density-density response via
the relation $\epsilon=1 -v\chi$.

For the HEG (or jellium) model, which is the most important reference for
simple metals, very accurate $f_{xc}$ kernels had been developed \cite{CP,CSOP,RA} by
fulfilling important exact constraints
(e.g. compressibility and the third-frequency-moment sum rules) and
using diffusion Monte Carlo input.

Here we extend those parametrization to the case of the JGM model.
In this Letter we restrict to the static  $\omega=0$ case and work in the adiabatic approximation.
In the JGM model {at density $n$ and with energy gap $E_g$},
the RPA static dielectric constant (i.e. with $f_{xc} = 0$) is \cite{LL}:
\begin{equation}
\epsilon_{0}^\mathrm{JGM-RPA}(q\to 0;n,E_g)=
1+\frac{4\pi n}{E_g^2}.
\label{e14}
\end{equation}
Eq. (\ref{e14}) is qualitatively different from the HEG counterpart,
being {\it finite} for a non-zero band gap. Similar models are widely 
used to describe dielectric properties of semiconductors \cite{gpp}.

In analogy to Eqs.  (20)-(21) of Ref. \cite{long_short_sottile}, a static kernel
for the JGM kernel ($f^\mathrm{JGM}_{xc}$) is approximated as:
\begin{equation}
f^\mathrm{JGM}_{xc}(q\to 0;n,E_g) \simeq
-\frac{4\pi}{q^2}\frac{1}{(\epsilon_0^\mathrm{JGM-RPA}
-1)}=-\frac{E_g^2}{n q^2}.
\label{e16}
\end{equation}
 We then extend the kernel of Ref. \cite{CP} to the 
JGM imposing condition (\ref{e16}) obtaining:
\begin{equation}
f_{xc}^\mathrm{JGM}(q;n,E_g)=\frac{4\pi}{q^2}B'(n,E_g)[e^{-k'_{n,E_g}q^2}-1]-\frac{4\pi}{k_F^2}
\frac{C'(n,E_g)}{1+1/q^2},
\label{e_Kernel}
\end{equation}
with
\begin{eqnarray}
B'(n,E_g) &=&\frac{B(n)+E_g}{1+E_g}, \;\;\; C'(n,E_g)= \frac{C(n)}{1+E_g}, \label{e_B_ker}\\
k'_{n,E_g} &=&  k_n+ \frac{1}{4\pi q^2} \frac{E_g^2}{n B'(n,E_g)} \label{e_k_ker}
\end{eqnarray}
where $k_n$, $B(n)$ and $C(n)$ are defined in Ref.~\cite{CP}, being
constructed from exact HEG constraints ($k_n$ and $C(n)$) and from
HEG diffusion Monte Carlo data \cite{CSOP} ($B(n)$); $k_F$ is the Fermi wavevector.

The JGM kernel has been constructed in order to satisfies the
following properties:

$(i)$ $f_{xc}^\mathrm{JGM}(q;n,E_{g}=0) \equiv f_{xc}^\mathrm{HEG}(q;n)$, where $f_{xc}^\mathrm{HEG}(q,n)$
is the HEG static kernel defined by Eq. (12) of Ref. \cite{CP}, that is very accurate
for the HEG correlation energy per particle {\it at every} wave vector $q$,
and it recovers the accurate (exact) RPA long-range effective interaction;

$(ii)$ $f_{xc}^\mathrm{JGM}(q;n,E_g\rightarrow \infty)\rightarrow -v(q)$,
such that this kernel gives a vanishing correlation energy ($E_c$) in the limit of \textit{perfect}
insulators ($E_g\rightarrow\infty$). Moreover, due to the
construction of $B'(n,E_g)$ and $C'(n,E_g)$, we observe by applying
the so-called adiabatic-connection fluctuation-dissipation
theorem \cite{CP}, that $E_c \approx constant/E_g + O(E_g^{-2})$, in the limit
of large band gap. This is one of the most important exact constraints
for the JGM which was derived from 
perturbation theory \cite{RS};

$(iii)$  $\lim_{q\rightarrow0}f_{xc}^\mathrm{JGM}(q;n,E_g)\approx -\alpha^\mathrm{JGM}(n,E_{g}) / q^2$,
where
\begin{equation}
\alpha^\mathrm{JGM}(n,E_{g})= 4\pi B'(n,E_g)
  \left[ 1-e^{ - E_{g}^2 / 4\pi nB'(n,E_g)} \right],
\label{limitins_2}
\end{equation}
 Eq. (\ref{limitins_2}) approaches to $E_g^2/n$ (see Eq. (\ref{e16})) for small $E_g$ (semiconductor case). 
Thus, in the $q\rightarrow0$ optical limit, we recover an LRC $\alpha/q^2$ behaviour;

$(iv)$ $\lim_{q\rightarrow\infty}f_{xc}^\mathrm{JGM}(q;n,E_g)=-\frac{4\pi}{k_F^2} C'(n,E_g)$.
This limit is unknown in the JGM, with the exception of the cases
$E_g\rightarrow 0$ and $E_g\rightarrow \infty$, for which the behaviour
of our kernel becomes exact. We recall that both RPA and ALDA fail badly in this limit \cite{CP}.

%
%

In order to use the JGM kernel Eq.~(\ref{e_Kernel})
 for real inhomogeneous bulk systems, we first consider its real space form $f_{xc}^\mathrm{JGM}(|\R-\R'|;n,E_g)$. 
We then use the DFT density $n(\R)$ in place of the constant density $n$,
whereas as a first approximation (valid for bulk systems investigated in this work) we approximate $E_{g}$ to the
fundamental band gap of materials; thus the kernel in the real space depends explicitly 
on the position (i.e. $f_{xc}^\mathrm{JGM}(|\R-\R'|,\R;E_g)$).
Then, we obtain the Fourier transform in the reciprocal space with the shape 
$\tilde{f}^\mathrm{JGM}_{xc}(|\mathbf{q} + \mathbf{G}|,\mathbf{G}-\mathbf{G}'; E_g)$,
with $\mathbf{q}$ vector in the first Brillouin zone, and $\mathbf{G}$, $\mathbf{G}'$
reciprocal lattice vectors. 
The kernel  still needs for symmetrization in $\mathbf{G},\mathbf{G}'$.
In previous works \cite{Valerio_thesisII,NLDA} the symmetrization was done in the real space.
In analogy to the symmetrization adopted for the
Coulombian term of the dielectric matrix \cite{Arya}, we perform the symmetrization in reciprocal space:
\begin{equation}
f_{xc}(\mathbf{q},\mathbf{G},\mathbf{G}'; E_g ) =
\tilde{f}^\mathrm{JGM}_{xc}(\sqrt{|\mathbf{q} + \mathbf{G}||\mathbf{q} + \mathbf{G'}|},\mathbf{G}-\mathbf{G}'; 
E_g).
\label{our_kernel_2}
\end{equation}
Eq. (\ref{our_kernel_2}) shows the following properties:

$(a)$ it reduces to Eq. (\ref{e_Kernel}) in the homogeneous case.

$(b)$ The head, $\mathbf{G}$=$\mathbf{G}'$=0, element of the kernel matrix, in the optical limit approaches 
$f_{xc}(q\rightarrow0,0,0;E_g) \approx - \langle\alpha\rangle q^{-2}$
where
\begin{equation}
\langle\alpha\rangle \equiv
  \int \alpha^\mathrm{JGM}(n(\mathbf{r}),E_{g}) d\mathbf{r},
\label{mean_value}
\end{equation}
which differs from $\alpha^\mathrm{JGM}$ evaluated with the average density $\bar{n}$.
Thus, an explicit value for the LRC $\alpha$ parameter is provided 
as the mean value of $\alpha^\mathrm{JGM}$  in the unit cell.  $\langle\alpha\rangle$ depends on the bandgap and it is \textit{density-functional} ( i.e.  it depends on the density inhomogeneity). 

$(c)$ In the $q\to 0$ limit, the wings of the kernel matrix
 $f_{xc}(q,0,\mathbf{G}'\neq 0;E_g)=f_{xc}(q,\mathbf{G}\neq 0,0;E_g)\propto 1/|\mathbf{q} + \mathbf{G}||\mathbf{q} + \mathbf{G'}|$ 
are $O(1/q)$ and the remaining elements are regular \cite{Kim_goerling,Gonze_Godby}. 

$(d)$ The ALDA-HEG terms are present in the kernel and they can play a role only at finite $q$, i.e.\ in energy-loss function calculations, where ALDA is already a good approximation \cite{EELS-ALDA}.

$(e)$ The computational cost is very low, comparable to standard ALDA.

The JGM kernel and its extension to inhomogeneous periodic systems (Eq. \ref{our_kernel_2}), is based on 
several well defined and physical approximations which need
to be tested in real situations. As a first test we applied Eq. (\ref{our_kernel_2}) to 
calculate the optical absorption of several 
bulk systems. 

The absorption spectrum is calculated as the $q \to 0$ limit of the imaginary part of the macroscopic dielectric function,
$\Im \epsilon^{M}(q,\omega)$, defined as
$\epsilon^{M}(q,\omega)=1/\epsilon^{-1}_{G=G'=0}(q,\omega)$.
%
We have implemented the $f_{xc}$ kernel in the linear response TD-DFT \texttt{dp} code 
\cite{dp-code}. The DFT ground state calculations are performed utilizing \texttt{ABINIT} \cite{MPmesh}.
Like in all previous approaches \cite{LRC,Botti_MgO,Bootstrap},
here we focus only on the e-h interaction, $f_{xc}^{e-h}$, excitonic effects part of the full $f_{xc}$,
while we simulate the effect of the e-e, $f_{xc}^{e-e}$, self-energy effects part by calculating 
$\chi_0$ with a scissor operator (so) bandgap corrected DFT-LDA electronic structure,
$E^{so}  = E_g - E_g^\mathrm{LDA}$,
where $E_g$ is set to the experimental band gap \cite{E_g_values}.
The same parameter $E_g$ is used into the JGM kernel.
Instead of the experimental band gap, the GW one can be used as well, so that to refer to first principle calculations.
Nevertheless, the problem to find the $f_{xc}^{e-e}$ kernel able to provide self-energy effects and the right $E_g$ (which is also an input of the JGM kernel), is once again postponed.
Hereafter, the RPA results are meant to be scissor operator shifted.

%
%
%
%
%
%
%
\begin{figure}
\includegraphics[width=\columnwidth]{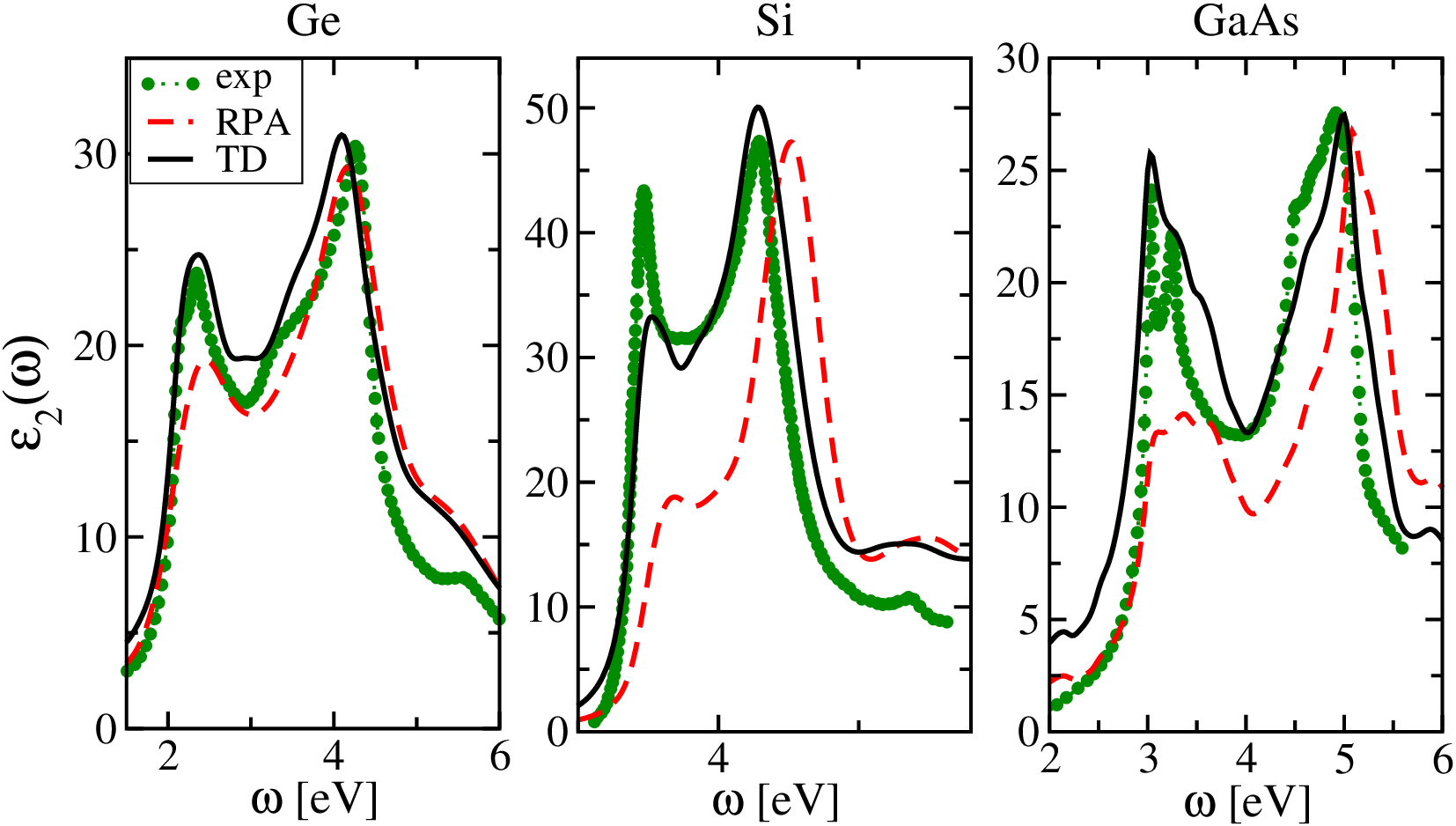}
\caption{\label{Ge_Si_GaAs} (color online). Imaginary part of the macroscopic dielectric function
($\epsilon_2(\omega)$)  for bulk Ge, Si and, GaAs. Red dashed line: RPA results with scissor operator
shifting. Black solid line:
TD-DFT with \textit{JGM} kernel. Green dots are the experimental absorption spectra: Ge from ref.
\cite{Ge}, Si from ref. \cite{Si}, GaAs from ref. \cite{GaAs}. Same notation has been utilized in the next Figures. Ge and Si are obtained with
relative Gaussian and Lorentzian broadening 0.02. Absolute Gaussian broadening 0.1 eV was used for GaAs.}
\end{figure}
%

In Fig. \ref{Ge_Si_GaAs} we show our calculated spectra for semiconductors: Ge, Si and GaAs.
These materials present weakly bound excitons  nearby the onset of the continuum.
Previous works \cite{Botti_MgO} have shown that e-h excitonic effects increase the oscillator strength of the lowest energy part of spectra.

Ge is the smallest band gap bulk system which we considered. 
 Excitonic effects
are modest, and the RPA calculation (dashed line) is already in good agreement with the experiment (green dots). 
Nevertheless, the TD-DFT JGM kernel (solid line) introduces improvements, enhancing the oscillator strength of the first peak at 2.2 eV.

On the other hand, in Si  e-h interactions have important effects on the
spectral weight of excitations and they determine a different absorption shape with respect to 
RPA. 
The JGM kernel increases the oscillator strength of the first feature at 3.4 eV where the RPA shoulder becomes a well defined peak, like in the experiment.

Similarly to Si, in GaAs excitonic effects shift the spectral weight to lower energies. The JGM kernel moves the peak near 5 eV towards the experimental position and it increases the intensity of the structure at 3 eV therefore achieving 
a very good agreement with the experiment.
%
%
%
\begin{figure}
\includegraphics[width=\columnwidth]{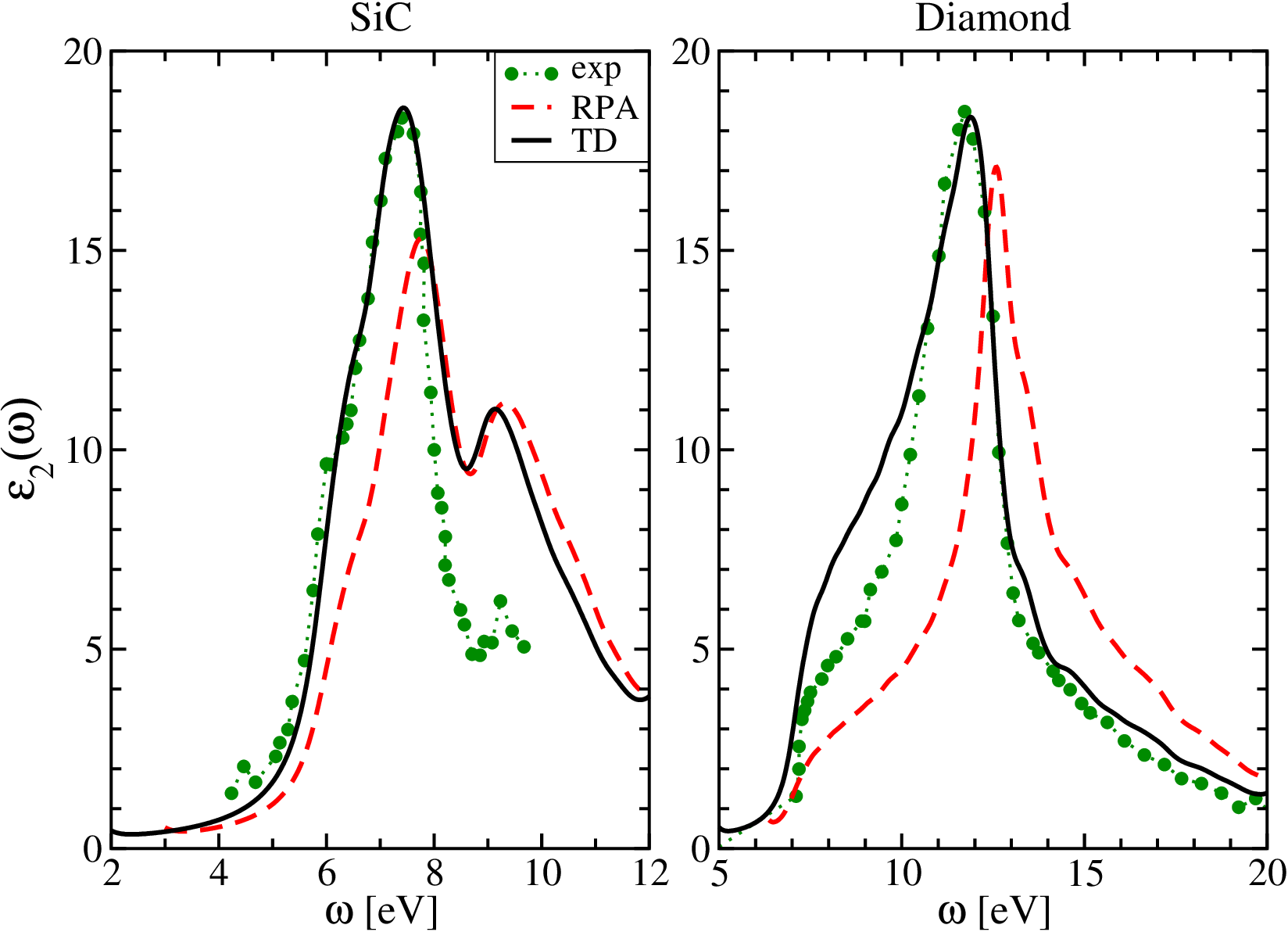}
\caption{\label{SiC_Diamond} (color online).
As in Fig. 1, but for SiC and Diamond bulk.
Experimental absorption spectra: SiC from Ref.
\cite{SiC} and Diamond from Ref. \cite{Diamond}. Absorption spectra are obtained with absolute
Lorentzian broadening: SiC: 0.3 eV, and Diamond 0.2 eV.}

\end{figure}
%

More severe tests are Diamond and SiC (Fig. \ref{SiC_Diamond}) where the larger band gap $E_g$ increases
the JGM kernel strength of excitonic effects. 
In SiC, the RPA peak at 8 eV is enhanced and red shifted in the TD-DFT result, achieving an almost perfect overlap with the experimental peak. The higher energy peak at 9 eV is reproduced at the correct experimental position, though is quite overestimated.

In Diamond, the position of the main RPA peak is correctly red shifted by the TD-DFT,
overlapping the  experimental position,
whereas the left shoulder is slightly too intensive. Overall, TD-DFT calculations show
a smooth and realistic description of optical spectra of both SiC and Diamond. 
%
\begin{figure}
\includegraphics[width=\columnwidth]{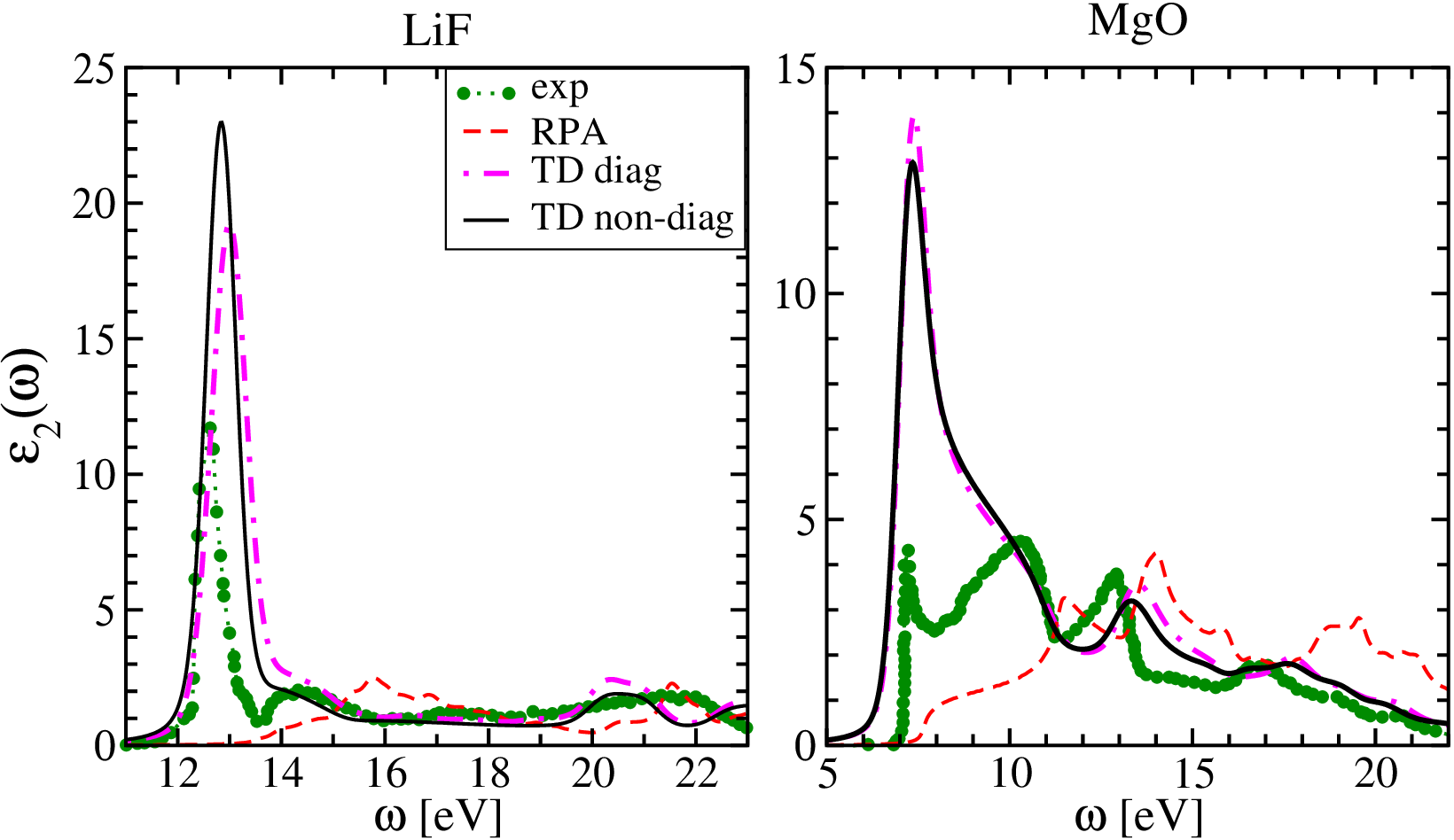}
\caption{\label{LiF_MgO} (color online).
As in Fig. 1, but for LiF and MgO.
Experimental absorption spectra: LiF from Ref.
\cite{LiF}, and MgO from Ref. \cite{Bechstedt}.
Here dotted-dashed magenta line is JGM TD-DFT only with diagonal $f_{xc}$.  
Absorption spectra obtained with the
absolute Gaussian broadening of 0.3 eV for both LiF and MgO.}
\end{figure}
%

LiF and MgO are wide-gap insulators, with a strongly bound exciton occurring  in
the band gap  \cite{Marini,Botti_MgO}.
They represent a severe workbench for any TD-DFT kernel.
In Fig. \ref{LiF_MgO}, we show the calculated absorption spectra for LiF and MgO in comparison with
the experiment. Here we report the results for both the full matrix JGM kernel (solid line), and also its $\mathbf{G}=\mathbf{G'}$ diagonal part only (dot-dashed), in order to show the effect of non-diagonal elements.
The RPA calculations completely miss both the main peaks
(LiF: $12.6$ eV and MgO: $7.2$ eV) and the resonant state features at higher energy.

In LiF, the diagonal TD-DFT JGM kernel is able to conjure 
the main excitonic peak, albeit it is blue shifted by 0.5
eV with respect to the experiment and the intensity is overestimated.
The second experimental peak at 14.5 eV is well captured
as shoulder, and it is joined in its final part.
High energy features at about 20 eV are present,
whereas the spurious BSE peak at 21 eV \cite{Marini} here is absent.
When inserting nondiagonal kernel elements, we have further excitonic strength 
that red shifts the main peak up to 12.80 eV, improving the agreement with the experiment.
Frequency dependence of $f_{xc}$ has been considered crucial for the correct description of the absorption spectra of
wide band gap insulators where the strongly bound excitons are appearing \cite{Marini,Botti_LRC}.
However, as shown in Ref. \cite{Bootstrap} and in this Letter, a static non-local kernel can also correctly address the e-h interactions in solid-state systems.

In MgO, the strongly bound exciton is accompanied by
high energy resonant peaks. 
JGM TD-DFT results display the strongly bound
excitonic peak at 7.2 eV in agreement with the experimental findings. Likewise the LiF case, the height
of the exciton is very intensive. The similarity with LiF concerns with the subsequent peak at 10.3 eV: in JGM TD-DFT calculations,
this becomes an artefact shoulder which joins the experimental results in the final part of the peak.
The overall behaviour at high energies is in rather good agreement with experiments: both 
energy and intensity peaks are well described. Non diagonal elements in the JGM kernel
slightly decrease the intensity of the main peak at 7.2 eV, while the 13.50 eV peak  is red shifted
to 13.30 eV in better agreement with the 13 eV experimental result.

\begin{table}
\caption{\label{tab1}
The fundamental band gap $E_g$ \cite{E_g_values}, 
$\alpha_{fit}$ from the best LRC fits,  
and  $\langle\alpha\rangle$ of Eq. (11).
}
\begin{ruledtabular}
\begin{tabular}{cccccc}
 &$E_g$ (eV)
 & $\alpha_{fit}\footnotemark[1]$
 & $\alpha_{fit}\footnotemark[2]$
 & $\langle\alpha\rangle$ \\
 \hline
Ge& 0.74 &  0.08 & ---  & 0.06 \\
Si& 1.17 &  0.20 & 0.13  & 0.11  \\
GaAs& 1.42 &  0.22 &0.15  & 0.22 \\
SiC& 2.36 & 0.50 & 0.20  & 0.23 \\
C  & 5.48 &  0.60 & 0.28  & 0.68 \\
MgO& 7.67 &  1.8 & ---  & 3.33  \\
LiF& 14.2 &  2.15 & 1.5  & 7.76  \\
\end{tabular}
\end{ruledtabular}
\footnotetext[1]{Ref. \cite{Botti_MgO}(for LiF was estimated by formula
$\alpha=4.615/\varepsilon^{stat}-0.213$)}
\footnotetext[2]{Ref. \cite{Botti_LRC}}
\end{table}
%

Finally, we stress that the most important contribution
to the absorption spectra, is given by the head of
the matrix kernel, that in the case
of a static kernel, as the one presented in this work, is just
an LRC-type  $\alpha/q^2$, and all our results can be well reproduced by this simple (non-empirical) LRC-type kernel.
In Table \ref{tab1}, we show a comparison between 
 JGM $\langle\alpha\rangle$ values  and  previous $\alpha$ estimations:
Results are well in agreement each other. 
This agreement begins to deviate for wider band gap bulk systems where the inhomogeneity 
of the density enhances, resulting in much higher values of $\langle\alpha\rangle$ for MgO and LiF.

In conclusion, we have found that  a simple, non-empirical and physical, static xc kernel
derived from the jellium-with-gap model (JGM) provides accurate optical absorption
spectra of a wide range of bulk solids (from metals to strong insulators).
The physics of this kernel relies on the validity of JGM, in spite of HEG, to represent real solids within TD-DFT.
This kernel can be easily implemented in any solid-state code, and
with a very reduced  computational cost. 
In this work we tested bulk systems, but JGM kernel can be also applied
to more complex systems, such as heterostructures or metal-organic interfaces, where the energy gap 
itself will depend on the position. The JGM kernel will correctly modulate the LRC contribution according to the local value of the gap. 
Further investigation will be devoted to energy-loss spectra at finite $q$.

This work was partially funded by the ERC Starting Grant FP7 Project DEDOM (No. 207441).
%
%
%


\begin{thebibliography}{[Vo]}
\bibitem{E_g_values}
The used $E_g$ values are (eV) taken from ref. \cite{E_g_values_ref} except for LiF \cite{E_g_LiF} and MgO \cite{Bechstedt}.
\end{thebibliography}
\end{document}